\documentclass[showpacs,pre]{revtex4}
\usepackage{amsmath}
\usepackage{graphicx}
\begin{document}
\title{Cooperative dynamics in coupled systems of fast and slow phase oscillators}
\author{Hidetsugu Sakaguchi and Takayuki Okita}
\affiliation{Department of Applied Science for Electronics and Materials,
Interdisciplinary Graduate School of Engineering Sciences, Kyushu
University, Kasuga, Fukuoka 816-8580, Japan}
\begin{abstract}
We propose a coupled system of fast and slow phase oscillators. We observe two-step transitions to quasi-periodic motions by direct numerical simulations of this coupled oscillator system. A low-dimensional equation for order parameters is derived using the Ott-Antonsen ansatz. The applicability of the ansatz is checked by the comparison of numerical results of the coupled oscillator system and the reduced low-dimensional equation. 
We investigate further several interesting phenomena in which mutual interactions between the fast and slow oscillators play an essential role. Fast oscillations appear intermittently as a result of excitatory interactions with slow oscillators in a certain parameter range. Slow oscillators experience an oscillator-death phenomenon owing to their interaction with fast oscillators. This oscillator death is explained as a result of saddle-node bifurcation in a simple phase equation obtained using the temporal average of the fast oscillations. Finally we show macroscopic synchronization of the order 1:m between the slow and fast oscillators.
\end{abstract}
\maketitle
\section{Introduction}
Coupled limit-cycle oscillators have been studied in various research fields such as physics, chemistry, biology, mechanical and electric engineering. In particular, mutual synchronization plays an important role in circadian rhythms, 
heartbeats, brain waves, etc. There has been intensive theoretical study of mutual synchronization in a large number of limit-cycle oscillators~\cite{rf:1,rf:2}. The Kuramoto model is a simple model which exhibits a phase transition from a disordered state to a synchronized state~\cite{rf:1,rf:3,rf:4,rf:5}. Ott and Antonsen developed a method to reduce the Kuramoto model to a low-dimensional dynamical system~\cite{rf:6}. Since then, many authors have studied Kuramoto type models using the Ott-Antonsen ansatz. Martens et al. studied coupled systems of two groups of oscillators with different average frequencies using the ansatz~\cite{rf:7}. In their model, two groups of oscillators interact with phase coupling. However, the interaction expressed by the phase difference is not suitable for two oscillators with a large frequency difference. Activator-inhibitor coupling is more frequently observed in chemical reactions and neural systems. 

For example, pacemaker neurons for respiratory rhythm were found in the ventrolateral medullary region called the pre-B\"otzinger complex. Recently, the role of glial cells, called astrocytes, in respiratory rhythmogenes has been studied. Slow calcium oscillation of astroglial cells was found in the pre-B\"otzinger complex~\cite{rf:8}. It was observed that rhythmic calcium elevation of astrocytes precedes the firing of neurons~\cite{rf:9}. Therefore, coupling between neurons and astrocytes might play an important role in rhythm generation. Oku et al. proposed a coupled system of one fast oscillator and one slow oscillator as a simple model of neurons and astrocytes~\cite{rf:10}.    

Motivated by this observation, we propose a coupled system of a large number of fast and slow phase oscillators in ‡U. In ‡V, we study the coupled phase oscillator model and find two-step transitions to quasi-periodic motion by direct numerical simulations. In ‡W, we derive a low-dimensional equation for order parameters using the Ott-Antonsen ansatz, and reproduce the two-step transitions to quasi-periodic motion by the numerical simulation of the low-dimensional equation. The main objective in this paper is to report some interesting cooperative dynamics in this coupled system of fast and slow oscillators in which mutual interactions play an essential role. We show intermittent occurrence of fast oscillation in ‡X, an oscillator-death state for slows oscillators caused by interaction with fast oscillators in ‡Y, and macroscopic synchronization of the order 1:m in ‡Z.   

\section{Coupled systems of fast and slow phase oscillators}
The model equation is a coupled active rotator model~\cite{rf:11}  expressed as 
\begin{eqnarray}
\frac{d\phi_{1i}}{dt}&=&\omega_{01}+\delta \omega_{1i}-b_1\sin\phi_{1i}+\frac{K_1}{N}\sum_{j=1}^N\sin(\phi_{1j}-\phi_{1i})+g_1S_2,\;{\rm for}\; i=1.,2,\cdots,N\nonumber\\
\frac{d\phi_{2i}}{dt}&=&\omega_{02}+\delta\omega_{2i}-b_2\sin\phi_{2i}+\frac{K_2}{N}\sum_{j=1}^N\sin(\phi_{2j}-\phi_{2i})+g_2S_1,\;{\rm for}\; i=1.,2,\cdots,N,
\end{eqnarray}
where $\phi_{1i}$ and $\phi_{2i}$ denote the phases of the fast and slow oscillators, respectively; $\omega_{01}$ and $\omega_{02}$  (where $\omega_{01}>\omega_{02}$) are the average values of the natural frequencies of the fast and slow oscillators, respectively; $\delta \omega_{1i}$ and $\delta\omega_{2i}$ denote the deviation of the natural frequency from the two average values;  $K_1$ and $K_2$ are coupling constants of the phase coupling in each group; and $b_1$ and $b_2$ are parameters which control the excitability. For example, each element in the first group behaves as an oscillator when $\omega_{01}+\delta\omega_{1i}>b_1$. However, it becomes an excitable element and leads to a stable stationary state for $t\rightarrow \infty$ when $b_1>\omega_{01}+\delta\omega_{1i}$, if interaction terms are absent. 
Interactions between the fast and slow oscillators are expressed by the last terms, $g_1S_1$ and $g_2S_2$, where $S_1$ denotes a signal from the slow oscillators to the fast oscillators, and $S_2$ denotes a signal from the fast oscillators to the slow oscillators. We assume that $S_1$ and $S_2$ are expressed by some functions of the order parameter $A_1=(1/N)\sum_{j=1}^N{\rm e}^{i\phi_{1j}}$ and $A_2=(1/N)\sum_{j=1}^N{\rm e}^{i\phi_{2j}}$. In this paper, we propose a model expressed by $S_1=1-(1/N)\sum_{j=1}^N\sin\phi_{1j}=1-{\rm Im}A_1,S_2=1-(1/N)\sum_{j=1}^N\sin\phi_{2j}=1-{\rm Im}A_2$. 

Near the excitable element-oscillator transition, i.e. $b_1\simeq \omega_{01}+\delta\omega_{1i}$, $\phi_{1i}$ stays close to $\pi/2$ for a long time, that is, $1-\sin\phi_{1i}$ is close to 0. When the element is excited,  the phase rotation occurs through $3\pi/2$ where $1-\sin\phi_{1i}$ increases to 2. Since the temporal average of $S_1$ increases continuously from 0 at the transition from excitable dynamics to oscillatory dynamics, $S_1$ is a quantity representing the activity of the oscillator, similar to the pulse frequency of neurons. Because $S_1$ and $S_2$ are always positive, $g_1>0$  implies excitatory coupling from slow oscillators to fast oscillators and the excitatory coupling makes the fast oscillator even faster. On the other hand, $g_2<0$ denotes inhibitory coupling from fast oscillators to slow oscillators, and the inhibitory coupling makes the slow oscillators even slower. Different types of cooperative dynamics are observed for other combinations of signs of $g_1$ and $g_2$. 

We assume that  $\delta\omega_{1i}$ and $\delta\omega_{2i}$ obey the Lorentz distributions:
\[p_1(\delta\omega_1)=\frac{\gamma_1}{2\pi}\frac{1}{(\delta\omega_1)^2+\gamma_1^2},\, p_2(\delta\omega_2)=\frac{\gamma_2}{2\pi}\frac{1}{(\delta\omega_2)^2+\gamma_2^2}.\]
Parameters $\gamma_1$ and $\gamma_2$ express the width of the natural frequency distributions for the fast and slow oscillators, respectively. 
\section{Transition to  macroscopic quasi-periodic motion}
\begin{figure}[t]
\begin{center}
\includegraphics[height=4.cm]{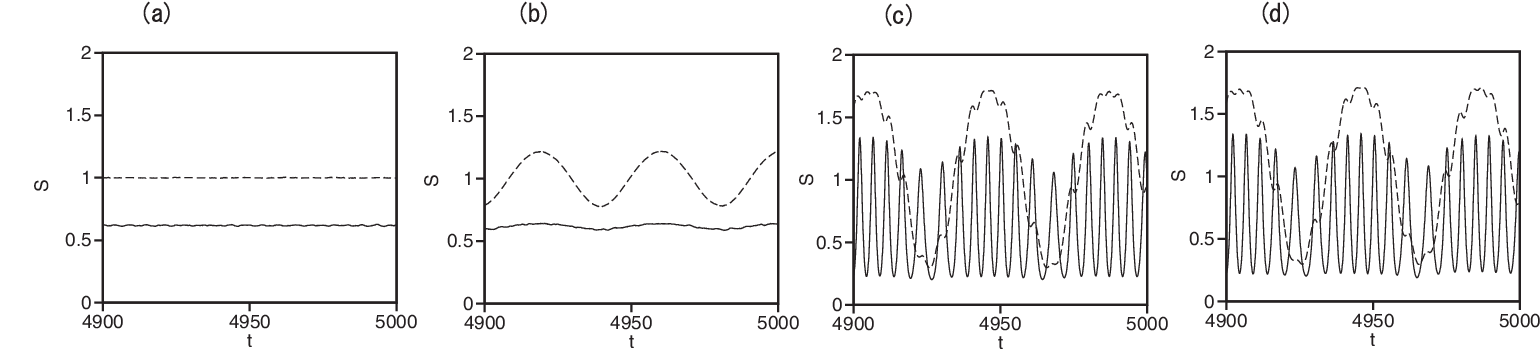}
\end{center}
\caption{Time evolutions of $S_1$ (solid line, representing fast oscillation) and  $S_2$ (dashed line, representing slow oscillation) obtained by the numerical simulation of  Eq.~(1) at (a) $K=0.015$, (b) 0.021 and (c) 0.04.  The number $N$ of oscillators is 5000. The initial condition is set to be $\phi_{1i}(0)=\phi_{2i}(0)=0$. (d) Time evolution of $S_1$ (solid line) and $S_2$ (dashed line) at $K=0.04$. The initial values of $\phi_{1i}$ and $\phi_{2i}$ are randomly distributed between 0 and $2\pi$.}\label{f1}
\end{figure}
\begin{figure}
\begin{center}
\includegraphics[height=4.cm]{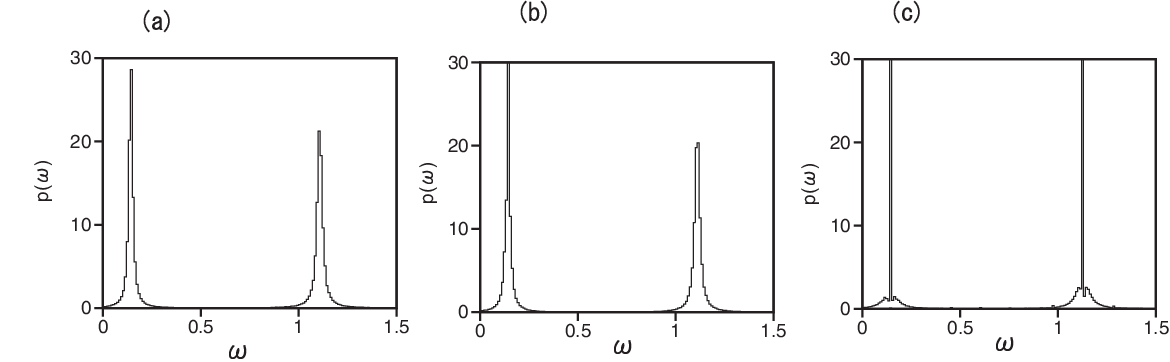}
\end{center}
\caption{Probability distributions of $p(\omega_1)$ (right)  and $p(\omega_2)$ (left) at (a) $K=0.015$, (b) $K=0.021$, and (c) $K=0.04$ obtained by the direct numerical simulations of Eq.~(1) with $N=5000$. Parts of $p(\omega)>30$ are cut. }
\label{f2}
\end{figure}
\begin{figure}
\begin{center}
\includegraphics[height=6.5cm]{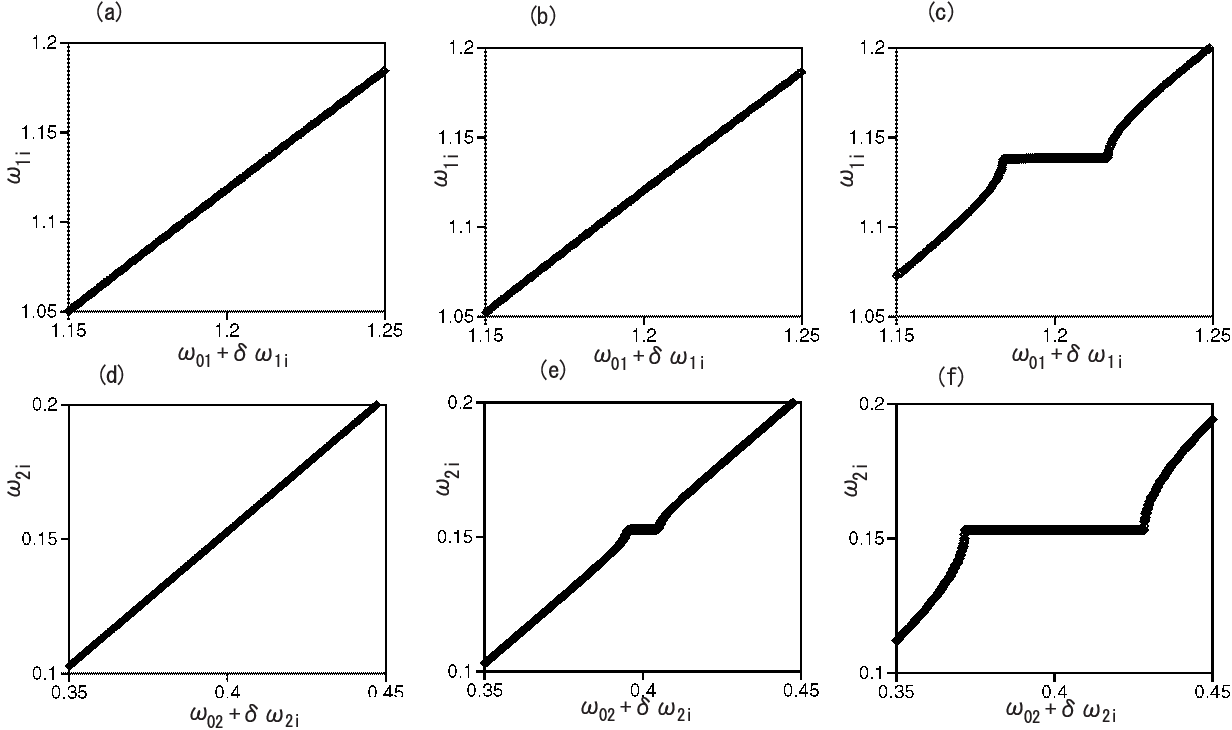}
\end{center}
\caption{Relationships between the natural frequency $\omega_{01}+\delta\omega_{1i}$ and $\omega_{1i}$ for fast oscillators at (a) $K=0.015$, (b) $K=0.021$, and (c) $K=0.04$. Relationships between the natural frequency $\omega_{02}+\delta\omega_{2i}$ and $\omega_{2i}$ for slow oscillators at (d) $K=0.015$, (e) $K=0.021$, and (f) $K=0.04$.}
\label{f3}
\end{figure}
In this section, we study transitions to quasi-periodic motion by direct numerical simulation of Eq.~(1). 
There are many parameters in our model system Eq.(1), and the whole parameter space is not surveyed yet. In the respiratory system, fast oscillators correspond to neurons, which exhibit pulsating (non-sinusoidal)  oscillation, and $b_1$ can take a nonzero value. Slow oscillators correspond to astrocytes which exhibit smooth sinusoidal oscillation, and parameter $b_2$ can be assumed to be zero. 
In this paper, we present some typical numerical results for $b_1=1$, $b_2=0$, assuming that $K_1=K_2$. In this section, parameters other than $K_1=K_2=K$ are fixed as $\omega_{01}=1.2,\omega_{02}=0.4,\,g_1=0.3,\,g_2=-0.4$, and $\gamma_1=\gamma_2=0.01$. The results are generic and qualitative behaviors do not change even if parameter values are varied slightly. 

Figures 1(a),(b) and (c) show the time evolutions of $S_1$ (solid line, representing fast oscillation) and  $S_2$ (dashed line, representing slow oscillation) obtained by the numerical simulation of  Eq.~(1) at (a) $K=0.015$, (b) 0.021 and (c) 0.04.  The number $N$ of oscillators is 5000. The initial condition is set to be $\phi_{1i}(0)=\phi_{2i}(0)=0$. A stationary state appears at $K=0.015$. 
Slow oscillation appears at $K=0.021$. A quasi-periodic motion is observed at $K=0.04$. In the quasi-periodic state, the fast oscillation of $S_1(t)$ is slowly modulated by the excitatory coupling with the slow oscillation of $S_2(t)$.  The slow oscillation of $S_2(t)$ is depressed by the inhibitory coupling with the fast oscillation of $S_1(t)$. Some fluctuations overlap on the  quasi-periodic motion. This could be because of the finite size effect of $N=5000$. 
The transition from the stationary state to the macroscopic oscillatory state occurs at the first critical point $K\sim 0.02$, which is a typical synchronization transition in a large population of slows oscillators. Mutual synchronization occurs among fast oscillators and a macroscopic quasiperiodic motion appears above the second critical point $K\sim 0.0256$.  The two-step transitions to the quasi-periodic motion are observed in wide parameter ranges. 
Figures 1(d) shows the time evolutions of $S_1$ (solid line) and  $S_2$ (dashed line) at $K=0.04$ for Eq.~(1) at the same set of parameters.  However, the initial values of $\phi_{1i}$ and $\phi_{2i}$ are randomly distributed between 0 and $2\pi$. Almost the same time evolution as Fig.~1(c) is observed, although the peak times of $S_1(t)$ and $S_2(t)$ are slightly different, i.e., there is a phase shift. Time evolutions of the order parameters do not depend on the initial conditions if the phase shift is neglected. This result suggests that there is a certain attractor in the macroscopic dynamics of order parameters.     

To investigate dynamical behavior of each oscillator, we have calculated the average frequency of each fast or slow oscillator as $\omega_{1i}=(\phi_{1i}(t_2)-\phi_{1i}(t_1))/(t_2-t_1),\, \omega_{2i}=(\phi_{2i}(t_2)-\phi_{2i}(t_1))/(t_2-t_1)$ for a large time interval $t_2-t_1$. 
Figures 2(a),(b), and (c) show the probability distributions $p(\omega_1)$ and $p(\omega_2)$ of the average frequency of fast and slow oscillators at (a) $K=0.015$, (b) $K=0.021$, and (c) $K=0.04$. These results are obtained by the direct numerical simulations of Eq.~(1) with $N=5000$. The probability distributions located around $\omega=1.1$ and $\omega=0.15$ are $p(\omega_1)$ and $p(\omega_2)$, respectively. 
At $K=0.015$, $p(\omega_2)$ has a form of Lorentz distribution.  
Sharp peaks like the $\delta$-function appear owing to the macroscopic synchronization at $K=0.04$. 
Figure 3(a),(b), and (c) show the relationship between the natural frequency $\omega_{01}+\delta\omega_{1i}$ and the time-average frequency $\omega_{1i}$ for fast oscillators at (a) $K=0.015$, (b) $K=0.021$, and (c) $K=0.04$. That is, Figs.~3(a),(b), and (c) are scatter plots of $(\omega_{01}+\delta\omega_{1i}, \omega_{1i})$ for $i=1,\cdots, N$ in a restricted range $1.15<\omega_{01}+\delta\omega_{1i}<1.25$. 
Figures 3(d),(e), and (f) show the relationship between the natural frequency $\omega_{02}+\delta\omega_{2i}$ and $\omega_{2i}$ for slow oscillators at (d) $K=0.015$, (e) $K=0.021$, and (f) $K=0.04$ in a restricted range $0.35<\omega_{02}+\delta\omega_{2i}<0.45$.
Flat regions imply mutual synchronization, which corresponds to the $\delta$-peak in the frequency distributions in Fig.~2. The ranges of $\omega_{01}+\delta\omega_{1i}$ and $\omega_{02}+\delta\omega_{2i}$ are restricted to show the flat regions clearly.  
Mutual synchronization of fast oscillators occurs at frequency $\omega_{1i}=1.14$ at $K=0.04$. Similarly, mutual synchronization of slow oscillators occurs at frequency $\omega_{2i}=0.153$ at $K=0.04$.  A small flat region appears in slow oscillators at $K=0.021$ in Fig.~3(e), however, there is no flat region  in fast oscillators at $K=0.021$ in Fig.~3(b). It is because the macroscopic synhcronization occurs for slow oscillators but does not occur for fast oscillators at $K=0.021$, that is, $K=0.021$ locates between the first and second critical points. 
\begin{figure}
\begin{center}
\includegraphics[height=4.cm]{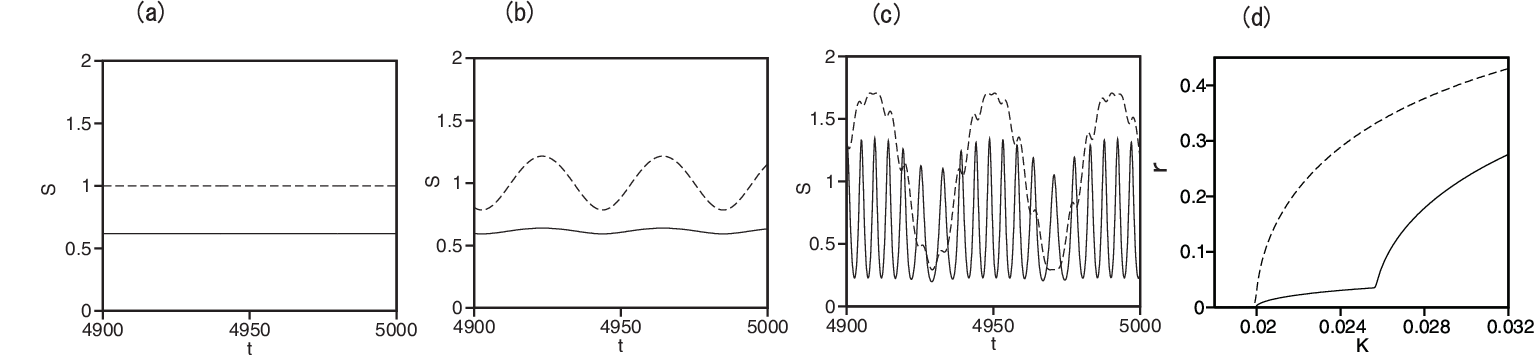}
\end{center}
\caption{Time evolutions of $S_1=1-$Im$A_1$ (solid curve) and $S_2=1-$Im$A_2$ (dashed curve) at (a) $K=0.015$, (b) 0.021, and $K=0.04$. (d) Magnitudes $r_1$ and $r_2$ of the oscillations as functions of $K$ calculated using Eq.~(4).}
\label{f4}
\end{figure}
\section{Low-dimensional equation for macroscopic motion}
In this section, we derive a low-dimensional equation for macroscopic motions using the Ott-Antonsen ansatz. 
The phase distributions for fast and slow oscillators with frequency deviation $\delta\omega_1$ and $\delta\omega_2$ are expressed $P_1(\phi_1,t,\delta\omega_1)$ and $P_2(\phi_2,t,\delta\omega_2)$, respectively. 
For the mean-field model expressed by Eq.~(1), the phase distribution functions  $P_1,P_2$ satisfy 
\begin{eqnarray}
\frac{\partial P_1}{\partial t}&=&-\frac{\partial}{\partial\phi_1}\left [\left \{\omega_{01}+\delta\omega_1+g_1S_2(t)+\frac{1}{2i}(H_1(t)e^{i\phi_1}-\bar{H}_1(t)e^{-i\phi_1})\right\}P_1\right ],\nonumber\\
\frac{\partial P_2}{\partial t}&=&-\frac{\partial}{\partial\phi_2}\left [\left \{\omega_{02}+\delta\omega_2+g_2S_1(t)+\frac{1}{2i}(H_2(t)e^{i\phi_2}-\bar{H}_2(t)e^{-i\phi_2})\right\}P_2\right ],
\end{eqnarray}
where $H_1(t)=-K_1\bar{A}_1(t)-b_1$ and $H_2=-K_2\bar{A}_2(t)-b_2$. Here, $\bar{A}$ is the complex conjugate of $A$. 
Assuming the Ott-Antonsen ansatz, the distributions $P_1(\phi_1)$ and $P_2(\phi_2)$ can be expanded as
\begin{eqnarray}
P_1(\phi_1,t,\delta\omega_1)&=&\frac{1}{2\pi}\left \{1+\sum_{m=1}^{\infty}\{a_1(t,\delta\omega_1)^me^{im\phi_1}+\bar{a}_1(t,\delta\omega_1)^me^{-im\phi_1}\}\right \},\nonumber\\
P_2(\phi_2,t,\delta\omega_2)&=&\frac{1}{2\pi}\left \{1+\sum_{m=1}^{\infty}\{a_2(t,\delta\omega_2)^me^{im\phi_2}+\bar{a}_2(t,\delta\omega_2)^me^{-im\phi_2}\}\right \},
\end{eqnarray}
where $\bar{a}_1(t,\delta\omega_1)$ and $\bar{a}_2(t,\delta\omega_1)$ denote the complex conjugates of $a_1(t,\delta\omega_1)$ and $a_2(t,\delta\omega_2)$. 

For the Lorentz distribution, the order parameters are expressed as $A_1(t)=\bar{a}_1(t,-i\gamma_1)$ and $A_2(t)=\bar{a}_2(t,-i\gamma_2)$. Then, the order parameters $A_1(t)$ and $A_2(t)$ obey the following coupled equations:
\begin{eqnarray}
\frac{dA_1}{dt}&=& (\mu_1+i\omega_{01}+ig_1S_2)A_1+(b_1/2)(1-A_1^2)-c_1|A_1|^2A_1,\nonumber\\
\frac{dA_2}{dt}&=& (\mu_2+i\omega_{02}+ig_2S_1)A_2+(b_2/2)(1-A_2^2)-c_2|A_2|^2A_2,
\end{eqnarray} 
where $\mu_1=K_1/2-\gamma_1,\mu_2=K_2/2-\gamma_2$, $c_1=K_1/2$, and $c_2=K_2/2$. Here, $S_1$ and $S_2$ are expressed as $S_1=1-{\rm Im} A_1$ and $S_2=1-{\rm Im} A_2$.
Equation (4) is rewritten using the phase and amplitude variables defined by  
 $\theta_1=\tan^{-1}({\rm Im} A_1/{\rm Re}A_1)$, $\theta_2=\tan^{-1}({\rm Im} A_2/{\rm Re}A_2)$, $R_1=|A_1|$, and $R_2=|A_2|$ as
\begin{eqnarray}
\frac{dR_1}{dt}&=&\mu_1R_1-c_1R_1^3+(b_1/2)(1-R_1^2)\cos\theta_1,\nonumber\\
\frac{d\theta_1}{dt}&=&\omega_{01}+g_1(1-R_2\sin\theta_2)-(b_1/2)(R_1+1/R_1)\sin\theta_1\nonumber\\
\frac{dR_2}{dt}&=&\mu_2R_2-c_2R_2^3+(b_2/2)(1-R_2^2)\cos\theta_2,\nonumber\\
\frac{d\theta_2}{dt}&=&\omega_{02}+g_2(1-R_1\sin\theta_1)-(b_2/2)(R_2+1/R_2)\sin\theta_2.
\end{eqnarray}
For $b_2=0$, $R_2=0$ for $\mu_2<0$, and $R_2=\sqrt{\mu_2/c_2}$ for $\mu_2>0$, and Eq.~(5) becomes coupled equations of the three variables $R_1,\theta_1$ and $\theta_2$. 

Thus, Eq.~(1) is reduced to a low-dimensional system Eq.~(4) or (5). 
We have performed numerical simulations of Eq.~(4) at various $K$'s  for $\omega_{01}=1.2,\omega_{02}=0.4,\,g_1=0.3,\,g_2=-0.4$, and $\gamma_1=\gamma_2=0.01$, which are parameter values used in the previous section.
The initial conditions are Re$A_1=0.1$, Im$A_1=0$, Re$A_2=0.1$, and Im$A_2=0$. 
Figures 4(a), (b), and (c) show time evolutions of $S_1=1-$Im$A_1$ (solid curve) and $S_2=1-$Im$A_2$ (dashed curve) at (a) $K=0.015$, (b) 0.021, and (c) $K=0.04$.
A stationary state is obtained at $K=0.015$, slow oscillation appears at $K=0.021$, and quasi-periodic motion is observed at $K=0.04$.  

It is noted that the time evolutions of $S_1$ and $S_2$ obtained by direct numerical simulations of Eq.~(1) shown in Fig.~1 is almost the same as the time evolutions of $S_1$ and $S_2$ calculated using the low-dimensional equations Eq.~(4) shown in Fig.~4, if phase shift and fluctuations by finite size effect are neglected. Although we have not yet proved mathematically that the invariant manifold corresponding to the Ott-Antonsen ansatz is a global attractor in the whole phase space, Eq.~(4) describes the dynamics of the order parameters very well, which strongly suggests that the Ott-Antonsen ansatz is applicable to our system.  

Detailed bifurcation structures can be investigated using Eq.~(4), because fluctuations by the finite-size effect do not appear and numerical simulations are much faster for Eq.~(4). Figure 4(d) shows the magnitudes $r_1$ and $r_2$ of the oscillations as functions of $K$ calculated using Eq.~(4).  Here, $r_1$ and $r_2$ are calculated as the root mean square of the time sequences of $S_1(t)$ and $S_2(t)$, respectively. For the sinusoidal oscillation, $r$ is equal to the amplitude of the oscillation divided by $\sqrt{2}$. The synchronization transition occurs  in the population of slow oscillators first at $K=K_c=2\gamma_2=0.02$ where $\mu_2=K/2-\gamma_2=0$ is satisfied. It is because the amplitude $R_2$ of macroscopic oscillations for the slow oscillators is zero for $\mu_2=K/2-\gamma_2<0$, and $R_2$ increases continuously from 0 for $\mu_2=K/2-\gamma_2>0$.  The quasi-periodic motion occurs at $K=0.0256$, where the magnitude $r_1$ increases rapidly. 

Stationary states, slow oscillation states, fast oscillation states, and quasi-periodic states are typical macroscopic states in our coupled systems of fast and slow oscillators. The four macroscopic states are observed in wide parameter ranges. In the following three sections, we discuss three topics of interesting cooperative dynamics in our coupled systems of fast and slow oscillators.  Mutual interactions expressed by $g_1$ and $g_2$ play an essential role in the cooperative dynamics. We investigate the coupled systems of fast and slow oscillators using numerical simulations of both Eq.~(1) and Eq.~(4).  In each case, we will check the applicability of the Ott-Antonsen ansatz by comparing numerical results of Eq.~(1) and (4).   
\begin{figure}
\begin{center}
\includegraphics[height=4.cm]{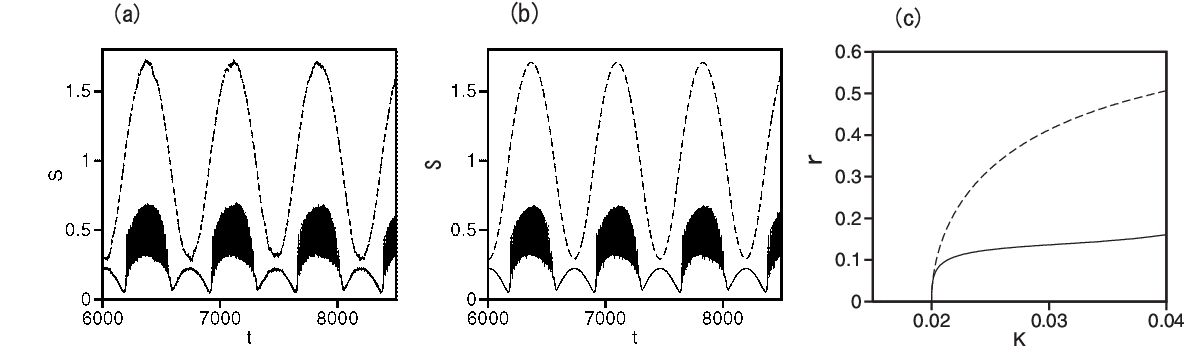}
\end{center}
\caption{(a) Time evolutions of $S_1(t)$ (solid curve) and $S_2(t)$ (dashed curve) at $K=0.04$ for $\omega_1=0.7,\omega_2=0.01, g_1=0.3$ and $g_2=-0.005$ in Eq.~(1). The initial condition is $\phi_{1i}(0)=\phi_{2i}(0)=0$. (b) Time evolutions of $S_1(t)$ (solid curve) and $S_2(t)$ (dashed curve) calculated using Eq.~(4) at the same parameters. (c) Magnitude $r_1$ (solid line) and $r_2$ (dashed line) of the oscillation as functions of $K$.}
\label{f5}
\end{figure}
\section{Waxing and waning dynamics}
When $\omega_{01}+\delta\omega_{1i}<b_1=1$, the oscillator of the first group becomes an excitable element when the coupling is absent. In this section, we discuss the waxing and waning behavior of fast oscillators. As a typical example, the parameters are fixed to be $\omega_1=0.7,\omega_2=0.01,g_1=0.3$, and $g_2=-0.005$.
Figure 5(a) shows the time evolutions of $S_1(t)$ (solid curve) and $S_2(t)$ (dashed curve) at $K=0.04$. The numerical simulation was performed using Eq.~(1) with $N=1000$. The initial condition is $\phi_{1i}(0)=\phi_{2i}(0)=0$. We have checked that almost the same time evolution is obtained in case that $\phi_{1i}(0)$ and $\phi_{2i}(0)$ take  random numbers between 0 and $2\pi$ as the initial condition.  
Figure 5(b) shows the time evolutions of $S_1(t)$ (solid curve) and $S_2(t)$ (dashed curve) using the  numerical simulation of Eq.~(4). These results show again that the time evolution of the order parameters does not depend on the initial values of $\phi_{1i}$ and $\phi_{2i}$ and the Ott-Antonsen ansatz is a good ansatz. 

Quasi-periodic motion appears in the time evolution of $S_1(t)$ similarly to the case of Figs.~1(c), 1(d), and 4(c). However, $S_1(t)$ exhibits a characteristic intermittent time evolution.  The originally excitable elements in the first group are excited by the slow oscillation and exhibit fast oscillation intermittently when $S_2(t)$ takes high values. On the other hand, the fast oscillation disappears when the level of $S_2(t)$ becomes low. That is, fast oscillations are induced by slow oscillations. This type of intermittent appearance of fast oscillations or the waxing and waning behavior is sometimes observed in neural systems. An example is spindle oscillation in brain waves, which appears during light sleep. The spindle oscillation is thought to appear in the thalamic network~\cite{rf:14}. 

Figure 5(c) shows the magnitudes $r_1$ and $r_2$ of the oscillation as functions of $K$ obtained using the numerical simulation of Eq.~(4). The macroscopic oscillation appears at $K=K_c=2\gamma_2=0.02$ which is the critical value of synchronization transition of slow oscillators. 
\section{Oscillator death of slow oscillators coupled with fast oscillators} 
In this section, we discuss an oscillator death state found in another parameter region. We vary $\gamma_1$ and $g_2$ as control parameters. Other parameters are fixed to be $\omega_{01}=1,\omega_{02}=0.1, K_1=K_2=0.025$, and $\gamma_2=0.01$.  The order parameter $R_2$ of the slow oscillation, $R_2=\sqrt{\mu_2/c_2}$, takes a positive value in Eq.~(5) at this parameter set. 
The macroscopic fast oscillation is expected to appear when $\gamma_1$ is small.
Figures 6(a)-(f) show the time evolutions of $S_1$ (solid line) and $S_2$ (dashed line) at (a) $\gamma_1=0.02,\,g_2=-0.25$, (b) $\gamma_1=0.02,\,g_2=-0.2$, (c) $\gamma_1=0.02,\,g_2=-0.145$, (d) $\gamma_1=0.006,\,g_2=-0.25$, (e) $\gamma_1=0.006,\,g_2=-0.2$, and (f) $\gamma_1=0.006,\,g_2=-0.145$ obtained by numerical simulations of Eq.~(1) with $N=1000$. The initial condition is $\phi_{1i}(0)=\phi_{2i}(0)=0$. Figures 7(a)-(f) show the time evolutions of $S_1$ (solid line) and $S_2$ (dashed line) obtained by numerical simulations of Eq.~(4) for the same parameter values as in Fig.~6.   Almost the same time evolutions are observed in the numerical simulation of Eq.~(1) and Eq.(4), although there are phase shifts and some fluctuations owing to the finite size effect overlap in the time evolutions shown in Fig.~4.  Slow oscillations are observed in Figs.~7(a) and (c), and quasi-periodic motions are observed in Figs.~7(d) and (f). A stationary state appears in Fig.~7(b). The stationary state corresponds to a stable stationary solution to Eq.~(5). Only fast oscillation appears in Fig.~7(e). Note that slow oscillations disappear even if $\mu_2>0$.
  
\begin{figure}
\begin{center}
\includegraphics[height=6.8cm]{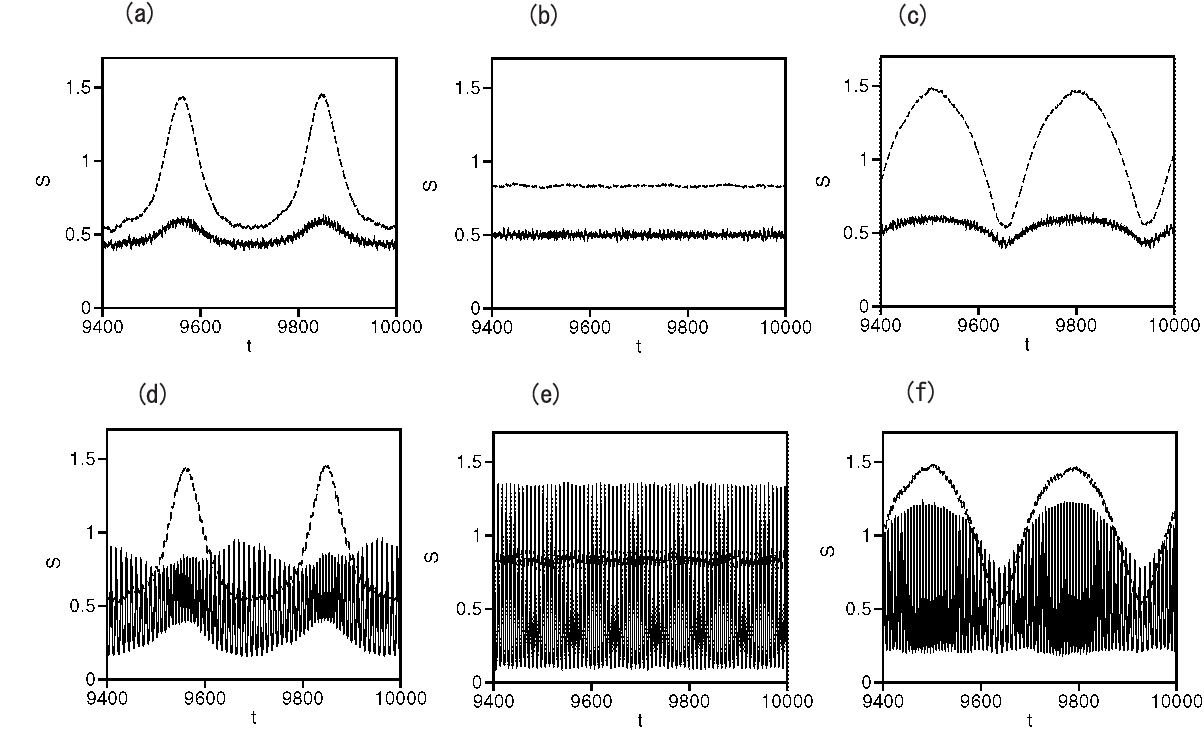}
\end{center}
\caption{Time evolutions of $S_1$ (solid line) and $S_2$ (dashed line) at (a) $\gamma_1=0.02,\,g_2=-0.25$, (b) $\gamma_1=0.02,\,g_2=-0.2$, (c) $\gamma_1=0.02,\,g_2=-0.145$, (d) $\gamma_1=0.006,\,g_2=-0.25$, (e) $\gamma_1=0.006,\,g_2=-0.2$, and (f) $\gamma_1=0.006,\,g_2=-0.145$ obtained by the numerical simulation of Eq.~(1). The initial condition is $\phi_{1i}(0)=\phi_{2i}(0)=0$. In Fig.~4-7, the other parameters values  are $\omega_{01}=1,\omega_{02}=0.1, K_1=K_2=0.025$, and $\gamma_2=0.01$.}
\label{f6}
\end{figure}
\begin{figure}
\begin{center}
\includegraphics[height=6.5cm]{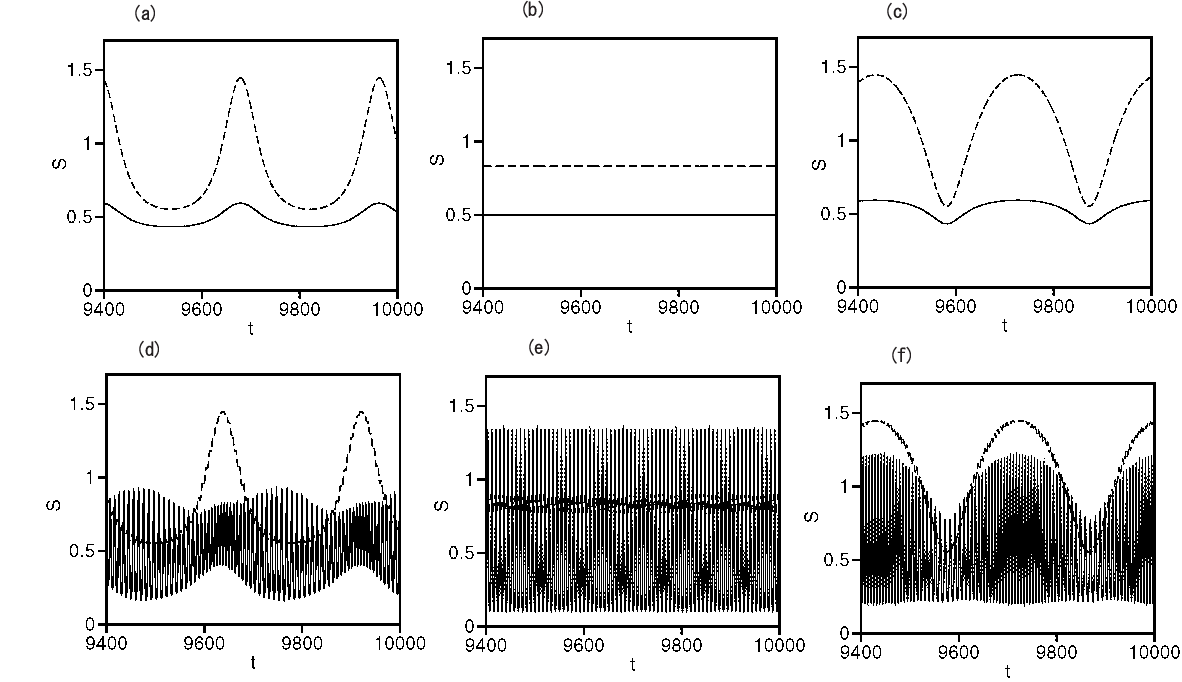}
\end{center}
\caption{Time evolutions of $S_1$ (solid line) and $S_2$ (dashed line) at (a) $\gamma_1=0.02,\,g_2=-0.25$, (b) $\gamma_1=0.02,\,g_2=-0.2$, (c) $\gamma_1=0.02,\,g_2=-0.145$, (d) $\gamma_1=0.006,\,g_2=-0.25$, (e) $\gamma_1=0.006,\,g_2=-0.2$, and (f) $\gamma_1=0.006,\,g_2=-0.145$ obtained by the numerical simulation of Eq.~(4).}
\label{f7}
\end{figure}
\begin{figure}
\begin{center}
\includegraphics[height=4.cm]{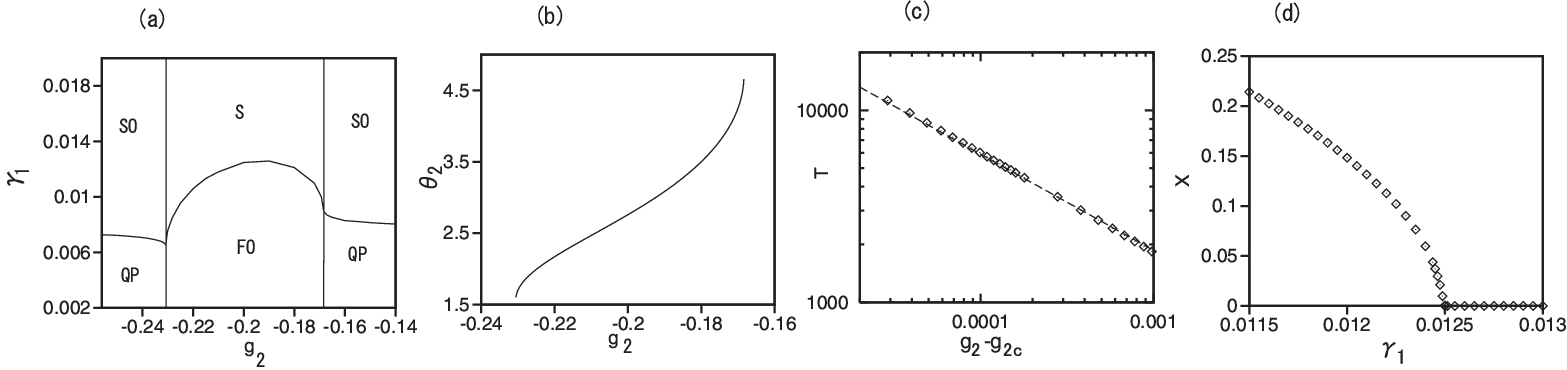}
\end{center}
\caption{(a) Phase diagram in a parameter space of $g_2$ and $\gamma_1$. 
Here, ``S" denotes the stationary state, ``SO"  slow oscillation, ``FO"  fast oscillation, and ``QP" quasi-periodic motion. (b) Phase $\theta_2$ of the slow oscillation as a function of $g_2$ at $\gamma_1=0.02$. (c) Period $T$ of the slow oscillations as a function of $g_2-g_{2c}$ near the saddle-node bifurcation point $g_{2c}=-0.1683$ at $\gamma_1=0.02$. (d) Maximum value of $X=$Re $A_1$ as a function of $\gamma_1$ at $g_2=-0.2$.}
\label{f8}
\end{figure}
\begin{figure}
\begin{center}
\includegraphics[height=4.cm]{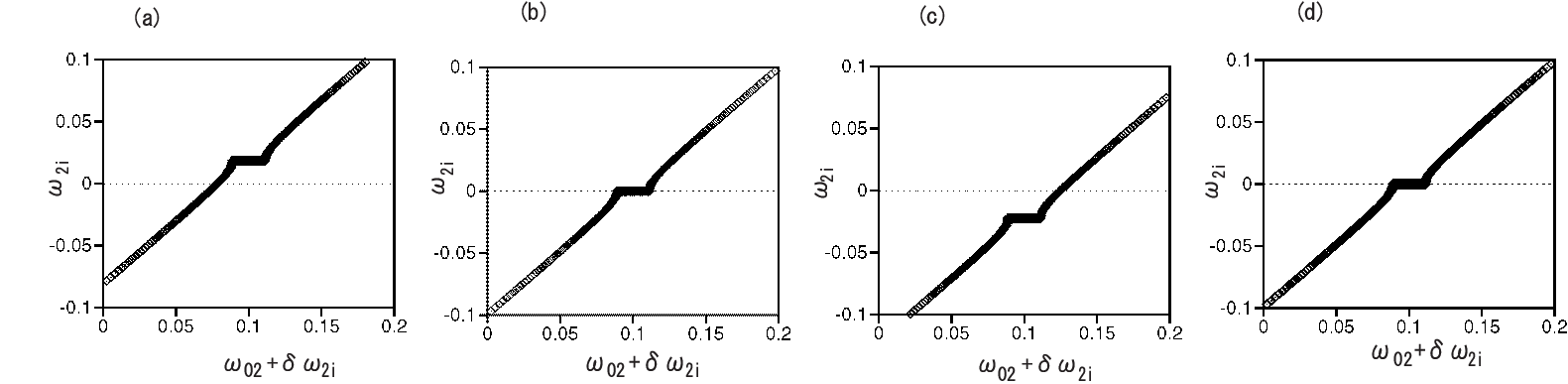}
\end{center}
\caption{Relationships between $\omega_{02}+\delta\omega_{2i}$ and $\omega_{2i}$ for (a) $g_2=-0.25,\,\gamma_1=0.02$, (b) $g_2=-0.2,\,\gamma_1=0.02$, (c) $g_2=-0.15,\,\gamma_1=0.02$, and (d) $g_2=-0.2,\,\gamma_1=0.006$ obtained using the numerical simulation of Eq.(1) with $N=1000$.}
\label{f9}
\end{figure}

Figure 8(a) shows a phase diagram in a parameter space of $g_2$ and $\gamma_1$ obtained by numerical simulations of Eq.~(4). There are four kinds of macroscopic states: stationary state, slow oscillation, fast oscillation, and quasi-periodic motion. They are  denoted by ``S", ``SO", ``FO", and ``QP", respectively.  This phase diagram is constructed by varying the parameters $g_2$ and $\gamma_1$ near the phase boundaries using Eq.~(4). The initial conditions are fixed to ${\rm Re}A_1=0.1,{\rm Im}A_1=0,{\rm Re}A_2=0.1,{\rm Im}A_2=0$ and the control parameters are changed stepwise. The macroscopic state is determined for each parameter set by the time evolutions of $S_1(t)$ and $S_2(t)$, and  observing that the attractor of Eq.~(4) is a fixed point, a limit-cycle, or a torus. 
We have performed numerical simulations using  several different initial values and obtained the same results.    
Figure 8(b) shows the phase $\theta_2$ of the slow oscillation in the stationary state as a function of $g_2$ for $\gamma_1=0.02$. The phase $\theta_2$ changes from $\pi/2$ to $3\pi/2$ as $g_2$ increases.  Transitions from the stationary state to slow oscillations occur at $g_2=g_{21c}\sim -0.1683$ and $g_2=g_{22c}\sim -0.2307$ for large values of $\gamma_1$. The transitions are saddle-node bifurcations, where $\sin\theta_2$ takes $\pm 1$ and stationary solutions disappear for $g<g_{22c}$ and $g>g_{21c}$. Figure 8(c) shows the period $T$ of slow oscillations as a function of $g_2-g_{21c}$ near the transition point $g=g_{21c}$ for $\gamma_1=0.02$ in a double logarithmic plot. 
The period increases as $T\sim 1/|g_2-g_{21c}|^{1/2}$ near the transition point, which is characteristic of the saddle-node bifurcation. 
The transition from the stationary state to the fast oscillation state is the supercritical Hopf bifurcation. 
Figure 8(d) shows the peak amplitude of the oscillation of ${\rm Re}A_1$ as a function of $\gamma_1$ for $g_2=-0.2$.  
The amplitude of the fast oscillation increases continuously from 0 at $\gamma_1=0.0125$. 
At the Hopf bifurcation, the stationary state changes into the fast oscillation state. Quasi-periodic motion appears at nearly vertical bifurcation lines of $g=g_{21c}\sim -0.1683$ and $g=g_{22c}\sim -0.2307$ from the fast oscillation state. Slow oscillation with a very long period overlaps with fast oscillation in the quasi-periodic state near the transition lines. This appears to be a kind of saddle-node bifurcation of the fast oscillation. The transitions from  slow oscillation to quasi-periodic motion occur near $\gamma_1\sim 0.007$ for $g_2<-0.2307$ and $\gamma_1\sim 0.0085$ for $g_2>-0.1683$. The transitions are the Hopf bifurcation of a limit-cycle corresponding to slow oscillation, or the bifurcation from a limit-cycle to a torus. This is interpreted as a synchronization-desynchronization transition of fast oscillators under the influence of slow oscillation. 

The information of each oscillator such as the average frequency is obtained by direct numerical simulation of Eq.~(1).  
Figures 9(a)-(d) show the relationship between $\omega_{02}+\delta\omega_{2i}$ and $\omega_{2i}$ for (a) $g_2=-0.25,\,\gamma_1=0.02$, (b) $g_2=-0.2,\,\gamma_1=0.02$, (c) $g_2=-0.15,\,\gamma_1=0.02$, and (d) $g_2=-0.2,\,\gamma_1=0.006$ calculated using Eq.~(1) with $N=1000$. 
Flat regions of $\omega_{2i}$ imply macroscopic mutual synchronization.  
The frequencies of the flat region in Figs.~9(a)-(d) are (a) $\omega=-0.0221$ (b) $\omega=0$ (c) $\omega=0.0184$, and (d) $\omega=0$. The entrainment frequency is negative for slow oscillations for $g_2<-0.2307$ and the entrainment frequency is positive for $g_2>-0.1683$. The entrainment frequency approaches 0 near the saddle-node bifurcation lines and becomes zero. The stationary state is different from the desynchonized state for $K/2<\gamma_2$ in which no flat region appears in the plot of  $\omega_2+\delta\omega_{2i}$ and $\omega_{2i}$.  The entrainment frequency of the slow oscillators is 0 even for the case of $\gamma_1<0.008$, where the fast oscillations appear as shown in Fig.~6(e). The macroscopic slow oscillation seems to be entrained to the zero frequency state for $-0.2307<g_2<-0.1683$. 

The bifurcation of the macroscopic entrainment of slow oscillation by the interaction with fast oscillation can be studied in greater detail for $\gamma_1=0$. 
At $\gamma_1=0$, $\mu_1=K/2=c_1$ and therefore $R_1=1$ in Eq.~(5). As the parameters $b_1$ and $b_2$ are set to be $b_1=1$ and $b_2=0$, Eq.~(5) is reduced to be
\begin{eqnarray}
\frac{d\theta_1}{dt}&=&\omega_1+g_1(1-R_2\sin\theta_2)-\sin\theta_1,\\
\frac{d\theta_2}{dt}&=&\omega_2+g_2(1-\sin\theta_1),
\end{eqnarray}
where $R_2=\sqrt{\mu_2/c_2}$. 
If the time evolution of $\theta_1$ is fast and $\theta_2(t)$ is sufficiently slow, then $\theta_2$ is assumed to be a constant in the time evolution of Eq.~(6). 
In this case, the probability distribution $P(\theta_1)$  is expressed as 
\begin{equation}
P(\theta_1)\propto \frac{1}{|d\theta_1/dt|}=\frac{1}{\omega_1+g_1(1-R_2\sin\theta_2)-\sin\theta_1}.
\end{equation}
The temporal average of $\sin\theta_1$ in Eq.~(7) with respect to the fast oscillation is evaluated using the probability distribution (8) as 
\begin{equation}
\langle \sin\theta_1\rangle=\frac{\int_{-\pi/2}^{\pi/2}\sin\theta_1P(\theta_1)d\theta_1}{\int_{-\pi/2}^{\pi/2}P(\theta_1)d\theta_1}=\omega_1+g_1(1-R_2\sin\theta_2)-\sqrt{\{\omega_1+g_1(1-R_2\sin\theta_2)\}^2-1}.
\end{equation}
The slowly varting component of $\theta_2(t)$ in Eq.~(7) therefore obeys
\begin{equation}
\frac{d\theta_2}{dt}=\omega_2+g_2[1-\omega_1-g_1(1-R_2\sin\theta_2)+\sqrt{\{\omega_1+g_1(1-R_2\sin\theta_2)\}^2-1}].
\end{equation}
Thus, $\sin\theta_2$ takes the value 1 at $g_2=g_{22c}$, and -1 at $g_2=g_{21c}$. 
The saddle-node bifurcations therefore occur at 
\begin{eqnarray}
g_{22c}&=&\frac{\omega_2}{\omega_1+g_1(1-R_2)-\sqrt{\{\omega_1+g_1(1-R_2)\}^2-1}-1},\nonumber\\
g_{21c}&=&\frac{\omega_2}{\omega_1+g_1(1+R_2)-\sqrt{\{\omega_1+g_1(1+R_2)\}^2-1}-1}.
\end{eqnarray}
The critical values are evaluated at $g_{22c}=-0.2307 $ and $g_{21c}=-0.1684$ for $\omega_{01}=1,\omega_{02}=0.1,g_1=0.3,K_1=K_2=0.025$, and $\gamma_2=0.01$. These values are consistent with  the numerical results. The macroscopic oscillations disappear between $g_{22c}$ and $g_{21c}$, and are locked to the zero-frequency state. This phenomenon is also interpreted as a kind of oscillator death of slow oscillators owing to the interaction with fast oscillators. This oscillator death is a macroscopic one, or the oscillator death of $S_2(t)$. In the level of individual oscillators, the mutually entrained slow oscillators stop oscillation in this state as shown in Figs.~9(b) and (d), while there are many desynchronized oscillators with nonzero average frequency.     
This type of oscillator death phenomenon is observed for other parameter sets approximately satisfying Eq.~(11). Other types of oscillator death phenomena have been studied in various coupled oscillator systems~\cite{rf:14,rf:15}.
\begin{figure}
\begin{center}
\includegraphics[height=6.5cm]{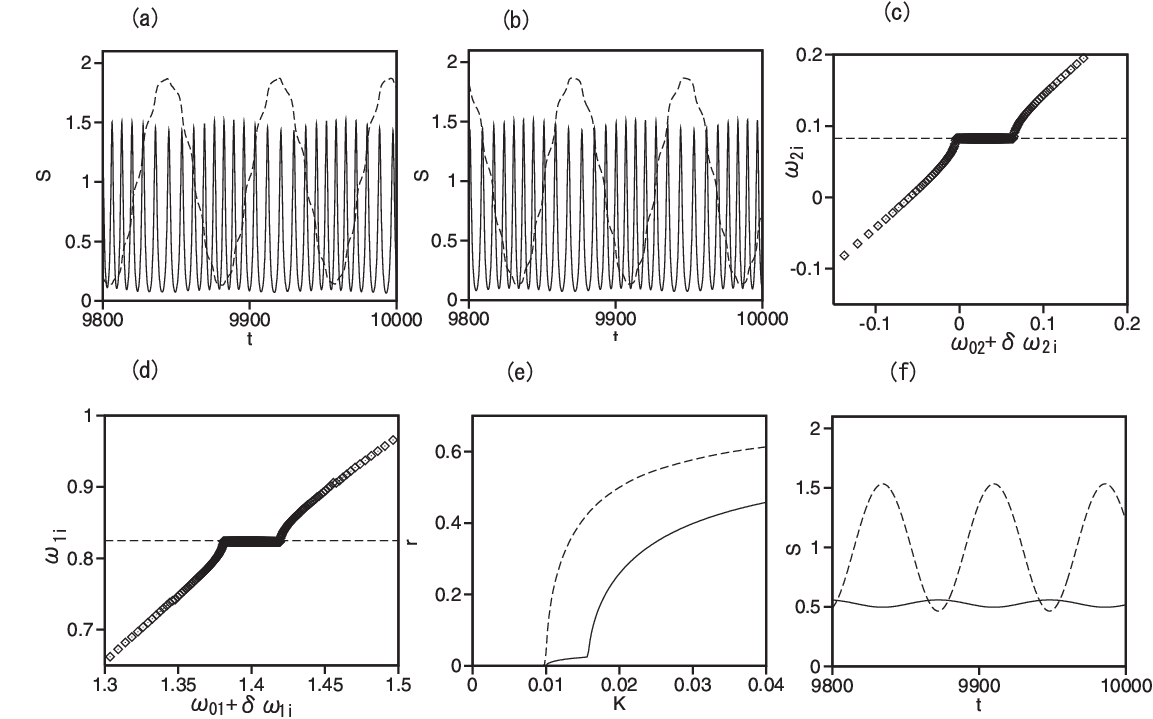}
\end{center}
\caption{(a) Time evolutions of $S_1(t)$ (solid line) and $S_2(t)$ (dashed line) obtained by numerical simulation of Eq.~(1) at $K=0.04$. Parameter values are set to be $\omega_{01}=1.4,\,\omega_{02}=0.03, g_1=-0.1,\,g_2=0.1$, and $\gamma_1=\gamma_2=0.005$. (b) Time evolutions of $S_1(t)$ (solid line) and $S_2(t)$ (dashed line) obtained by numerical simulation of Eq.~(4) at $K=0.04$. (c) Relationships between $\omega_{02}+\delta\omega_{2i}$ and $\omega_{2i}$. The horizontal line is $\omega_{2i}=0.0825$. (d) Relationships between $\omega_{01}+\delta\omega_{1i}$ and $\omega_{1i}$. The horizontal line is $\omega_{1i}=0.825$. (e) Magnitudes $r_1$ (solid line) and $r_2$ (dashed line) of the oscillation as a function of $K$. (f) Time evolutions of $S_1(t)$ and $S_2(t)$ obtained using the numerical simulation of Eq.~(4) at $K=0.014$.}
\label{f10}
\end{figure}
\section{Macroscopic synchronization of order 1:m}
In this section, we study the case of $g_1<0$ and $g_2>0$. In this case, mutual synchronization between fast oscillators and slow oscillators is often observed. Because there is a frequency difference between fast and slow oscillators, synchronization of the order $n:m$ occurs. In this section, we present numerical results for $\omega_{01}=1.4,\,\omega_{02}=0.03, g_1=-0.1,\,g_2=0.1$, and $\gamma_1=\gamma_2=0.005$. However, various types of locking were observed by varying parameters such as $\omega_{01}$ and $\omega_{02}$.  (A similar macroscopic entrainment between fast oscillation and slow oscillation is often observed in the case of $g_1<0$ and $g_2<0$.)

 Figure 10(a) shows time evolutions of $S_1(t)$ and $S_2(t)$ obtained by numerical simulation of Eq.~(1) with $N=1000$ at $K=0.04$. As an initial condition, $\phi_{1i}(0),\phi_{2i}(0)$ take random values between 0 and $2\pi$. Here, $S_1(t)$ exhibits ten pulsating oscillations in a period of $S_2(t)$, and $S_1(t)$ is slightly depressed when $S_2(t)$ reaches its peak value, because of the inhibitory interaction of $g_1<0$.  Figure 10(b) shows the time evolutions of $S_1(t)$ and $S_2(t)$ obtained by the numerical simulation of Eq.~(4) for the same parameter set. Figures 10(a) and (b) show again the applicability of the Ott-Antonsen ansatz. 

The average frequency of each oscillator can be calculated using Eq.~(1). 
Figure 10(c) shows a relationships between $\omega_{02}+\delta\omega_{2i}$ and $\omega_{2i}$ obtained by the numerical simulations of Eq.~(1) with $N=1000$. The horizontal dashed line is $\omega_{2i}=0.0825$, which represents the entrainment frequency. Figure 10(d) shows a relationship between  $\omega_{01}+\delta\omega_{1i}$ and $\omega_{1i}$. The horizontal dashed line is $\omega_{1i}=0.825$.  Figures 10(c) and (d) imply that the synchronization of the order 1:10 occurs. Ten pulsating oscillations appear for the entrained fast oscillators in the flat region of Fig.~10(d), after $S_2(t)$ reaches a peak value. That is, the macroscopic slow oscillations appear to play the role of leading the  fast oscillators. This may be related to the role of astrocytes in the respiratory system. 

Figure 10(e) shows the magnitudes $r_1$ (solid line) and $r_2$ (dashed line) of the oscillation as a function of $K$ calculated using Eq.~(4). The macroscopic slow oscillation appears at $K=2\gamma_2=0.01$, and the macroscopic fast oscillation appears at $K=0.016$. The macroscopic synchronization of the order 1:10 is observed for $K>0.016$. Figure 10(f) shows time evolutions of $S_1(t)$ and $S_2(t)$ obtained by the numerical simulation of Eq.~(4) at $K=0.014$.  It is clear that $S_1(t)$ exhibits slow oscillation as an effect of $S_2$ at $K=0.014<0.016$.

\section{Summary}
We studied synchronization-desynchronization transitions in coupled systems of fast and slow oscillators. We performed direct numerical simulation of the Kuramoto-type phase oscillator model and a reduced model derived using the Ott-Antonsen ansatz. Similar time evolutions were observed for numerical simulations of the Kuramoto-type model and the reduced model. 
The good agreement suggests that the Ott-Antonsen ansatz is applicable to our model system. 

We found various phenomena in this coupled system of fast and slow oscillators. A quasi-periodic motion appears as a result of two-step transitions.  In the case that $\omega_1$ is smaller than $b_1$, fast oscillation appears intermittently as a result of excitatory interaction with the slow oscillators, which is similar to the waxing and waning behavior. We also found a type of oscillator death phenomenon of slow oscillators due to the interaction with fast oscillators.
The oscillator death phenomenon is thought to appear as a result of the saddle-node bifurcation in the phase equation for slow oscillations obtained using the temporal average of the fast oscillations. Finally, we found a macroscopic synchronization of the order 1:m in case of $g_1<0$ and $g_2>0$. 

Various states, such as a stationary state, slow oscillations, fast oscillations, quasi-periodic motion, and a synchronized state of the order of 1:m appear in our coupled system of fast and slow oscillators.  The parameter sets used in this paper are somewhat restricted. However, these oscillatory states, and the transitions among the various states, are generic and are expected to appear in wide parameter ranges.

\end{document}